# Superconductivity above 50K in $Ln$FeAsO$_{1-y}$ ($Ln$= Nd, Sm, Gd, Tb and Dy) Synthesized by High-pressure Technique

Kiichi Miyazawa[1,2], Kunihiro Kihou[1], Parasharam M. Shirage[1], Chul-Ho Lee[1,3], Hijiri Kito[1,3], Hiroshi Eisaki[1,3], Akira Iyo[1,2,3]*

[1] National Institute of Advanced Industrial Science and Technology, Tsukuba, Ibaraki 305-8568, Japan

[2] Department of Applied Electronics, Tokyo University of Science, 2641 Yamazaki, Noda, Chiba 278-3510, Japan

[3] JST, Transformative Research-Project on Iron Pnictides (TRIP), 5, Sanbancho, Chiyoda, Tokyo 102-0075, Japan

**Abstract**

We have succeeded in synthesizing single-phase polycrystalline samples of oxygen-deficient oxypnictide superconductors, $Ln$FeAsO$_{1-y}$ ($Ln$: lanthanide elements) with $Ln$=La, Ce, Pr, Nd, Sm, Gd, Tb and Dy using high-pressure synthesis technique. It is found out that the synthesis pressure is the most important parameter for synthesizing single-phase samples, in particular for the heavier $Ln$'s, such as Tb and Dy. The lattice parameters systematically decrease with the atomic number of $Ln$, reflecting the shrinkage of $Ln$ ionic radius. For the lighter $Ln$'s (La, Ce, Pr, Nd), $T_c$ increases monotonously with decreasing the lattice parameters from 26K for $Ln$=La to 54K for $Ln$=Nd, then stays at the constant value around 53K for the heavier counterpart (Nd, Sm, Gd, Tb and Dy). The results suggest the intimate relationship between the crystal structural parameters and the superconductivity on the one hand, as well as the possible existence of the inherent maximum $T_c$ on the other, which is located around 50 K in the $Ln$FeAsO based materials.

E-mail: iyo-akira@aist.go.jp, kiichi-miyazawa@aist.go.jp
*Corresponding author

1. Introduction

Discovery of the superconductivity with a transition temperature ($T_c$) of 26 K in the Fe-based oxypnictide LaFeAsO$_{1-x}$F$_x$ has provided a new paradigm for exploring new high-$T_c$ superconductors.[1,2] Soon after the discovery, it was reported that the replacement of La by other rare-earth element ($Ln$), such as Ce, Sm, Pr, Nd and Gd, significantly increases $T_c$.[3,4] To date, the highest $T_c$ exceeds 50 K.[5-9] The structure of the parent compound $Ln$FeAsO contains alternate stacking of $Ln$O and FeAs layers along the $c$-axis. Superconductivity is induced by the partial substitution of fluorine (F) for oxygen (O) in the $Ln$O planes, which leads to the electron doping in FeAs planes. Instead of the F-substitution, Ren *et al*. and Kito *et al*. have synthesized the O-deficient $Ln$FeAsO$_{1-y}$ using high-pressure synthesis technique and found that the system also exhibits high-$T_c$ superconductivity,[10,11] with its $T_c$ comparable to the F-substituted counterpart.

Common to the F-substituted and the O-deficient cases, $T_c$ strongly depends on the variation of $Ln$. It is well known that $T_c$ of the La-based materials is only 26 K, almost the half of the highest $T_c$ recorded for $Ln$=Nd, Sm, and Gd, even though the fundamental crystal structure is unchanged. This fact suggests that the superconductivity in this system strongly depends on the subtle change of the lattice parameters, and have triggered a lot of experimental and theoretical studies.[12-15]

For the heavier $Ln$'s ($Ln$=Gd, Tb and Dy), the sample quality and their superconducting properties depend on the synthesis condition. In particular for $Ln$=Tb and Dy, superconducting samples can be synthesized only using high-pressure technique. Furthermore, to our knowledge, there is no report on the successful synthesis of single-phase Dy- and Tb-based superconductors with their $T_c$'s exceeding 50K.[16-18] Accordingly, it is not clear yet whether the lower $T_c$ of the heavier $Ln$-based samples is really intrinsic to the system, or due to the poorer sample quality. The relationship between $Ln$ and $T_c$ in $Ln$FeAsO based materials is therefore still open.

In this paper, we report the high-pressure synthesis of single-phase samples of $Ln$FeAsO$_{1-y}$ ($Ln$=La, Ce, Pr, Nd, Sm, Gd, Tb and Dy). We have found that the optimization of synthesis pressure is the key for obtaining single-phase samples, which is particularly effective for heavier $Ln$=Sm, Gd, Tb and Dy. Moreover, we have also found that $T_c$ increases with improving sample purity. Using the single-phase samples, we have established the $Ln$-$T_c$ relationship for the $Ln$FeAsO based superconductors, which shows that the $Ln$FeAsO$_{1-y}$ ($Ln$=Nd, Sm, Gd, Tb and Dy) inherently possesses $T_c$ higher than 50K, irrespective of the variation of $Ln$.

## 2. Experimental

In this study, polycrystalline samples were prepared by the high-pressure synthesis method using a cubic-anvil-type apparatus (Riken CAP-07). As (chip), Fe (coarse powder), Fe$_2$O$_3$ (powder), $Ln$ (chip) (all purities ≥ 99.9%) are used as the starting materials. First, $Ln$As was synthesized by reacting $Ln$ and As chips at 500 °C for 15 hours, and then 850°C – 950°C for $Ln$=La, Ce, Pr, Nd and Sm, or 1000 °C for $Ln$=Gd, Tb and Dy for 5 hours in an evacuated quartz tube. The purity of $Ln$As phase was confirmed every time by the powder X-ray diffraction (XRD) analysis. $Ln$As, Fe and Fe$_2$O$_3$ were mixed at nominal compositions of $Ln$FeAsO$_{1-y}$ ($y$=0.60 - 0.70) and ground by an agate mortar in a glove box filled with dry nitrogen gas. The mixed powder was pressed into a pellet. The pellet was heated in a BN crucible at about 1050-1150°C under a pressure of about 2.0 to 5.5 GPa for 2 h. The synthesis temperature was 1050-1100°C for $Ln$=Gd, Tb, Dy, and 1100-1150°C for $Ln$=La, Ce, Pr, Nd and Sm. As described later, the synthesis pressure plays an essential role in synthesizing single-phase samples, in particular for *Ln*=Gd, Tb and Dy.

Powder XRD patterns were measured using CuK$_a$ radiation (Rigaku Ultima IV). The dc magnetic susceptibility was measured using a SQUID magnetometer (Quantum Design MPMS). First, the sample was cooled down to 5 K without applying a magnetic field. Then after a magnetic field of 5 Oe was applied, the zero-field cooling (ZFC) curve was measured by warming the sample up to above $T_c$. Then, the field-cooling (FC) curve was measured by cooling the sample down to 5 K. The resistivity was measured by a standard four-probe method.

## 3. Results and Discussion

*3.1 Effect of the synthesis pressure*

Fig. 1 (a) and (b) show the powder XRD patterns of the samples with a nominal composition DyFeAsO$_{0.7}$ synthesized at various pressures ranging from 2.0 to 5.5 GPa. As indicated in Fig. 1 (a), extra peaks due to the DyAs impurity phase are observed in the samples synthesized at pressures of 2.0 and 3.5 GPa, which apparently go away with increasing the synthesis pressure. Similar tendency is observed also for *Ln*=Tb, Gd and Sm, in which increasing the synthesis pressure up to 5.0 GPa significantly improves the sample purity.

With increasing the synthesis pressure, the diffraction peaks systematically shift toward the higher angle, indicating the shrinkage of the lattice parameters. To highlight the shift, the (212) peaks are shown in an extended scale in Fig. 1(b). Based on the XRD pattern, the lattice parameters (*a*- and *c*-axis in a tetragonal form) are calculated and plotted in Fig. 2. The *a*- and $c$-axis parameters shrink by about 0.4 and 0.5 %, respectively, when the synthesis pressure increases from 2.0 to 5.5 GPa.

Fig. 3 shows the temperature ($T$-) dependence of the magnetic susceptibility of DyFeAsO$_{0.7}$ synthesized at various pressures. The sample synthesized at 2.0 GPa (not shown) does not show a superconducting transition down to 5 K. The samples synthesized at 3.5, 4.5 and 5.5 GPa show superconductivity with their $T_c$'s increasing with increasing pressure, from 37 K (3.5GPa), 44 K (4.5GPa), eventually up to 51 K (5.5 GPa), as indicated by the arrows in Fig. 3. The transition width also sharpens with increasing synthesis pressure, indicating the improved sample homogeneity. The volume fractions estimated from the magnitude of diamagnetic signal at 5 K (ZFC) are about 53 % (3.5 GPa), 61 % (4.5 GPa) and 63 % (5.5 GPa).

Fig. 4 represents the relationship between $T_c$ and the $a$-axis lattice parameter of $Ln$=Sm, Gd, Tb and Dy samples synthesized under various pressures ranging from 2.0 to 5.5 GPa. It is immediately noticed that the lattice parameters strongly depend on the synthesis pressure, and also that there is a strong correlation between the lattice parameters and $T_c$. Increase of the synthesis pressure results in the decrease of the lattice parameter, along with the increase of $T_c$, especially for $Ln$=Tb and Dy. For all $Ln$'s, the highest $T_c$'s which exceed 50K are recorded for the samples synthesized under high pressures, typically 5.0 - 5.5 GPa. (Susceptibility and resistivity data are shown in Fig. 7 and Fig. 8).

For $Ln$FeAsO$_{1-y}$ with light $Ln$'s, it is known that superconductivity appears concomitantly with the shrinkage of lattice parameters.[19] In this case, the amount of O-deficiency ($y$) is the controlling parameter and superconductivity is induced by introducing $y$. Also for the heavier $Ln$'s, it is likely that the primary effect of higher-pressure synthesis to stabilize the O-deficiency phase which is more difficult to be formed compared to the lighter

*Ln*'s. Neutron diffraction analysis will be helpful to clarify the effect of the synthesis pressure on the crystal structure and oxygen content of the samples.

*3.2 Ln dependence of lattice parameters*

Fig. 5 shows the powder XRD patterns of the *Ln*FeAsO$_{1-y}$ samples for *Ln*=La, Ce, Pr, Nd, Sm, Gd, Tb and Dy. The current XRD pattern indicates that samples are free from impurity phases and all of the major peaks can be indexed assuming the *Ln*FeAsO structure with *P4/nmm* symmetry. In particular, as already shown in Fig. 1, the peaks due to *Ln*As which often appear in the poorly synthesized samples are completely gone in the current sample series. The diffraction peaks systematically shift toward the higher angles with increasing atomic number (i.e. with decreasing ionic radius) of *Ln*. The lattice parameters changes almost linearly with the change of the *Ln*$^{3+}$ ionic radius,[20] as shown in Fig. 6. The lattice parameters are summarized together with $T_c$ in Table I.

*3.3 Superconducting transition temperature*

Fig. 7 (a) to (h) show the *T* - dependence of the magnetic susceptibility of the samples we have obtained so far for *Ln*FeAsO$_{1-y}$ (*Ln*=La, Ce, Pr, Nd, Sm, Gd, Tb and Dy). Here we show the data of the highest $T_c$ samples for each *Ln* series, which means that the data presented in Fig. 7 provide the *lower boundaries* of the possibly-attainable highest-$T_c$'s for the *Ln*FeAsO$_{1-y}$ superconductors, not to say that they are the real highest $T_c$'s. The sharp drops of the magnetic susceptibility, corresponding to the onset of superconductivity, are observed for all the samples both in the ZFC and the FC curves. $T_c$ ($\chi$-onset) is determined from the intersection of the two extrapolated lines; one is drawn through the susceptibility curve in the normal state just above the transition, and the other is drawn through the susceptibility curve that clearly deviate from the normal state background, see the inset in Fig. 7 (a). $T_c$ ($\chi$-onset) are summarized in Table I. $T_c$ increases from $Ln$=La to Nd and apparently exceeds 50 K for $Ln$=Nd, Sm, Gd, Tb and Dy. The volume fractions estimated from the magnitude of diamagnetic signal at 5 K (ZFC) are about 46, 43, 31, 80, 60, 70, 64 and 63 % for $Ln$=La, Ce, Pr, Nd, Sm, Gd, Tb and Dy, respectively, thus ensuring the bulk superconductivity.

Figs. 8 (a) to (h) shows $T$- dependence of the resistivity for $Ln$=La, Ce, Pr, Nd, Sm, Gd, Tb and Dy, respectively. All the samples show metallic ($d\rho/dT > 0$) behavior in all $T$-range. As shown in the insets, sharp superconducting transitions are observed. $T_c$ ($\rho$-onset) is determined from the intersection of the two extrapolated lines; one is drawn through the resistivity curve in the normal state just above transition, and the other is drawn through the resistivity curve that clearly deviate from the normal state resistivity as shown in the insets in Fig. 8. $T_c$ ($\rho$-onset) and $T_c$ from the zero-resistivity, $T_c$ ($\rho$-zero), are summarized in Table I. $T_c$ ($\rho$-mid) is determined at the temperature where the resistivity is 50% of its value at the $T_c$ ($\rho$-onset). Again, one can see the same tendency as observed in the magnetic susceptibilities, namely, $T_c$ increases from $Ln$=La to Nd and exceeds 50 K for $Ln$=Nd, Sm, Gd, Tb and Dy.

In our samples, $T_c$ ($\chi$-onset) agrees well with $T_c$ ($\rho$-mid) or $T_c$ ($\rho$-onset) within $\pm 2$ K in most cases, ensuring that the superconducting properties in our samples are homogeneous. in comparison to the previous studies. Also in our samples, $T_c$ ($\rho$-zero), $T_c$ ($\rho$-mid) and $T_c$ ($\rho$-onset) are close to one another. It is worth mentioning that there are many papers reporting the record-breaking high $T_c$'s so far. However, their $T_c$'s are rather ill-defined in some cases, mainly due to the poor superconducting properties. For example, one can easily find the papers in which $T_c$ is defined by $T_c$ ($\rho$-onset), while $T_c$ ($\rho$-zero) or $T_c$ ($\chi$-onset) are located more than 10 K lower, which makes the $T_c$ values less trustable. Such uncertainly can be ruled out in our samples.

The $Ln$-dependence of the $T_c$ ($\chi$-onset) for the $Ln$FeAsO$_{1-y}$ samples is summarized in Fig. 9, which reveals the systematic tendency against the $a$-axis lattice parameter. $T_c$ increases from 28 K ($Ln$=La) to 53 K ($Ln$=Nd), almost linearly to the $a$-axis parameter. Once $T_c$ reaches 53 K, it stays almost constant from $Ln$=Nd to Dy. As mentioned, at this moment one cannot rule out the possibility for further increase of $T_c$ by finer tuning of the synthesis conditions. However, as far as the observed tendency is concerned, the current results appear to suggest the existence of an upper limit of $T_c$ for the $Ln$FeAsO$_{1-y}$ system, which is located at 53 K. Note that $T_c$ is constant for heavier $Ln$ even though the $a$- and $c$-parameters decrease with the atomic number. This implies that the lattice parameters themselves are not dominant factors for determining the maximum $T_c$ in the heavier side of $Ln$'s. .

## 4. Summary

Single-phase polycrystalline samples of oxygen-deficient oxypnictide superconductors, $Ln$FeAsO$_{1-y}$ with $Ln$=La, Ce, Pr, Nd, Sm, Gd, Tb and Dy have been prepared using high-pressure synthesis technique. It is found that the lattice parameters, $T_c$, the phase purity of the samples strongly depend on the synthesis pressure, in particular for heavier $Ln$'s, such as Tb and Dy. Sharp superconducting transitions were observed both in resistivity and magnetic susceptibility measurements. It is demonstrated that the $Ln$FeAsO$_{1-y}$ superconductors containing heavier $Ln$'s (Nd, Sm, Gd, Tb and Dy) inherently possesses high-$T_c$ exceeding 50 K.

**Acknowledgements**

The authors thank M. Ishikado, N. Takeshita, T. Ito, Y. Tomioka, Y. Tanaka, S. Kouno, K. Tokiwa, S. Mikusu, T. Watanabe for valuable discussions. This work was supported by Grant-in-Aid for Specially promoted Research (20001004) from The Ministry of Education, Culture, Sports, Science and Technology (MEXT) and JST, Transformative Research-Project on Iron Pnictides (TRIP).

**Figure captions**

Fig. 1 (a) Powder XRD patterns for the samples with a nominal composition of DyFeAsO$_{0.7}$ synthesized at various pressures ranging from 2.0 to 5.5 GPa. (b) The enlargement of (212) peaks around 2$\theta$ (deg.)=57~58 (b).

Fig. 2 Lattice parameters of *a*- and *c*-axis for the samples with a nominal composition of DyFeAsO$_{0.7}$ synthesized at various pressures ranging from 2.0 to 5.5 GPa.

Fig. 3 Temperature dependence of susceptibility for the samples with a nominal composition of DyFeAsO$_{0.7}$ synthesized at pressures of 3.5, 4.5 and 5.5 GPa. $T_c$'s are shown by arrows.

Fig. 4 Relationships between $T_c$ and *a*-axis lattice parameter for the samples with a nominal composition of *Ln*FeAsO$_{0.7}$ (*Ln*=Sm, Gd, Tb and Dy) synthesized under various pressures ranging from 2.0 to 5.5 GPa. $T_c$'s are determined by magnetic susceptibility measurements.

Fig. 5 Powder XRD patterns for the samples with a nominal composition of *Ln*FeAsO$_{0.7}$ (*Ln*=La, Ce, Pr, Nd, Sm, Gd, Tb and Dy). The synthesis pressure of the samples was 2.0 GPa for *Ln*=La, Ce, Pr, Nd, 5.0 GPa for *Ln*=Sm, Gd, Tb, and 5.5 GPa for *Ln*=Dy. Peaks are indexed assuming the *Ln*FeAsO structure with *P4/nmm* symmetry.

Fig. 6 Ionic radius of $Ln^{3+}$ (coordination number is 8) dependence of the *a*- and *c*-axis lattice parameters for the *Ln*FeAsO$_{1-y}$ samples (*Ln*=La, Ce, Pr, Nd, Sm, Gd, Tb and Dy).

Fig. 7 Temperature ($T$)- dependence of the magnetic susceptibility of the $Ln$FeAsO$_{1-y}$ samples for $Ln$=La, Ce, Pr, Nd, Sm, Gd, Tb and Dy. The inset in Fig. 7 (a) shows the definition of $T_c$ ($\chi$-onset). The synthesis pressure of the samples was 2.0 GPa for $Ln$=La, Ce, Pr, Nd, 5.0 GPa for $Ln$=Sm, Gd, Tb, and 5.5 GPa for $Ln$=Dy.

Fig. 8 Temperature ($T$)- dependence of the resistivity of the $Ln$FeAsO$_{1-y}$ samples for $Ln$=La, Ce, Pr, Nd, Sm, Gd, Tb and Dy. The insets show the $T$-dependence of resistivity near $T_c$ and the definition of $T_c$ ($\rho$-onset). The synthesis pressure of the samples was 2.0 GPa for $Ln$=La, Ce, Pr, Nd, 5.0 GPa for $Ln$=Sm, Gd, Tb, and 5.5 GPa for $Ln$=Dy.

Fig. 9 $Ln$-dependence of $T_c$ ($\chi$-onset) for the $Ln$FeAsO$_{1-y}$ samples ($Ln$=La, Ce, Pr, Nd, Sm, Gd, Tb and Dy). $T_c$ is plotted against the $a$-axis lattice parameter.

Table I. Lattice parameters of $a$- and $c$-axis and $T_c$ for the $Ln$FeAsO$_{1-y}$ samples ($Ln$=La, Ce, Pr, Nd, Sm, Gd, Tb and Dy).

| $Ln$ | $a$ (Å) | $c$ (Å) | $T_c$ ($\rho$-onset) | $T_c$ ($\rho$-mid) | $T_c$ ($\rho$-zero) | $T_c$ ($\chi$-onset) |
|---|---|---|---|---|---|---|
| La | 4.0257 | 8.7190 | 29.0 | 28.4 | 26.2 | 27.8 |
| Ce | 3.9950 | 8.6310 | 41.2 | 40.2 | 35.5 | 38.5 |
| Pr | 3.9627 | 8.5721 | 49.2 | 48.5 | 46.0 | 48.1 |
| Nd | 3.9432 | 8.5290 | 54.3 | 53.7 | 52.1 | 53.0 |
| Sm | 3.9218 | 8.4522 | 53.3 | 52.1 | 50.5 | 52.5 |
| Gd | 3.8905 | 8.3928 | 54.0 | 53.6 | 52.4 | 53.5 |
| Tb | 3.8753 | 8.3527 | 52.6 | 52.2 | 50.2 | 52.4 |
| Dy | 3.8625 | 8.3216 | 51.6 | 51.2 | 49.8 | 51.2 |

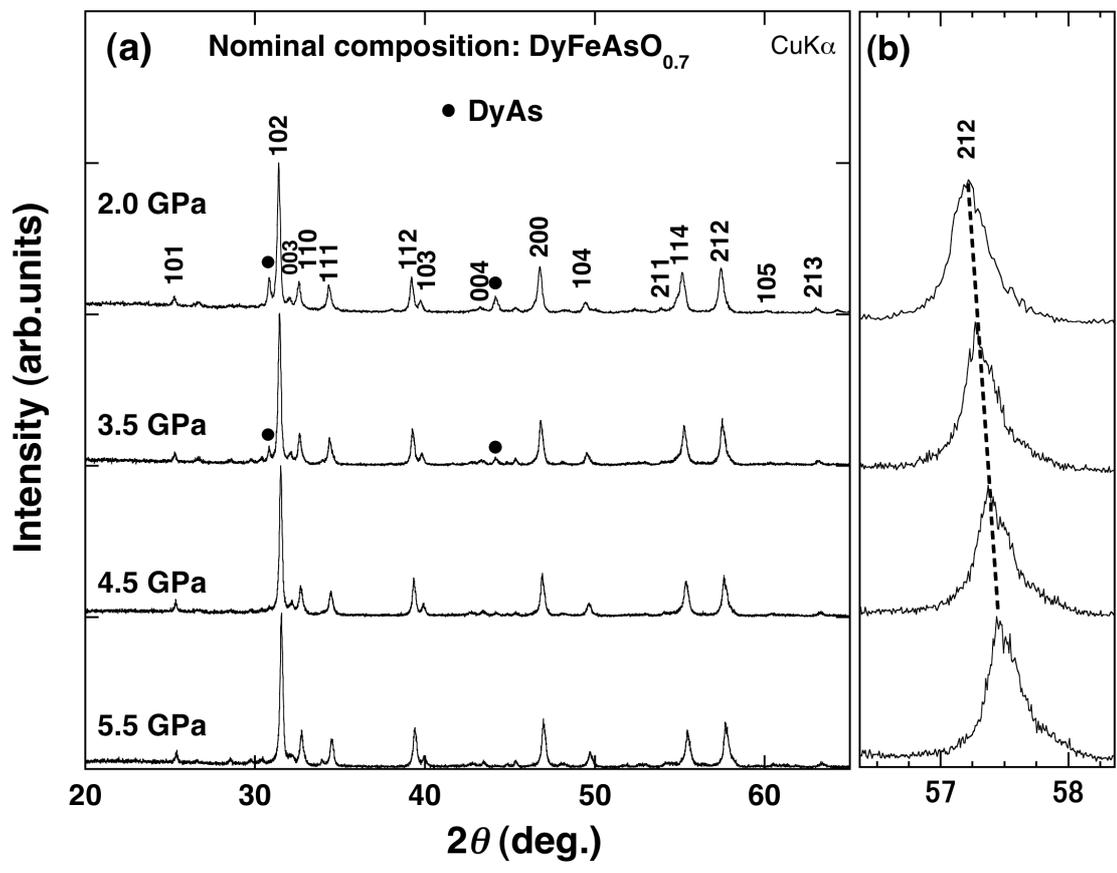

Fig. 1

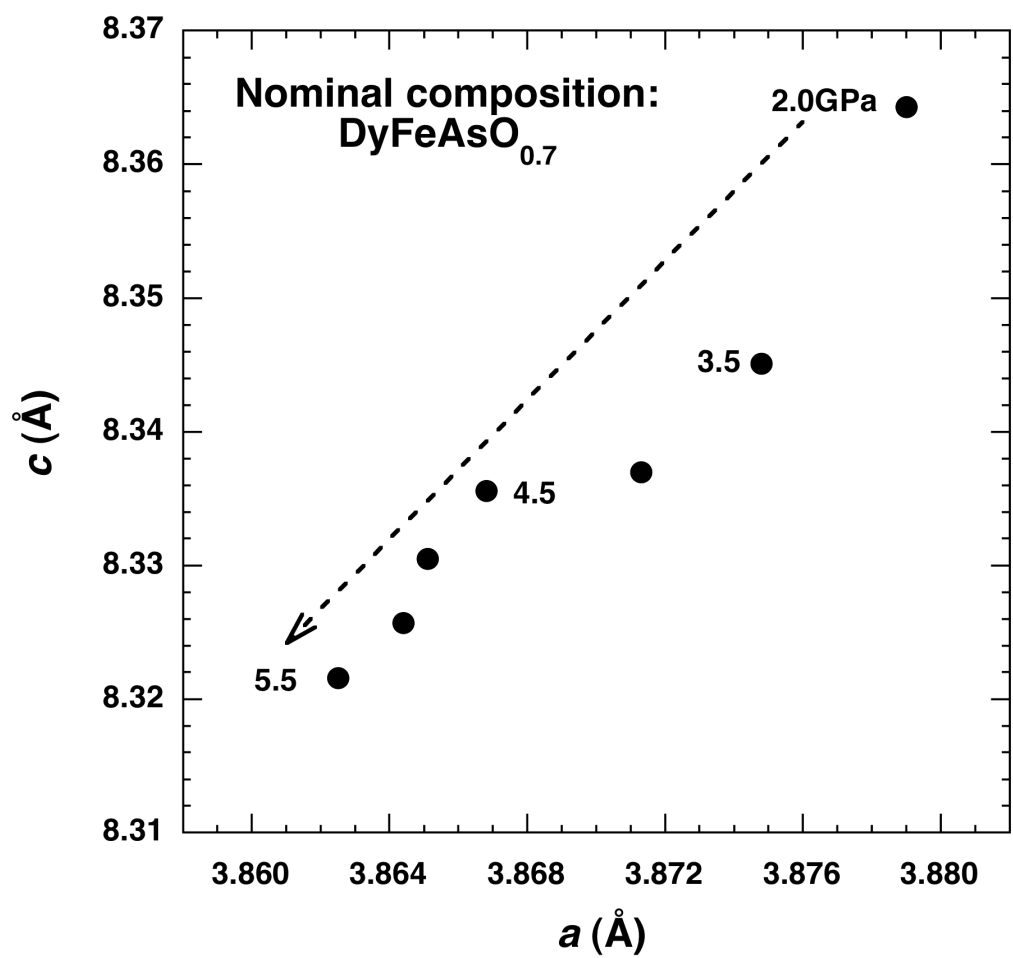

Fig. 2

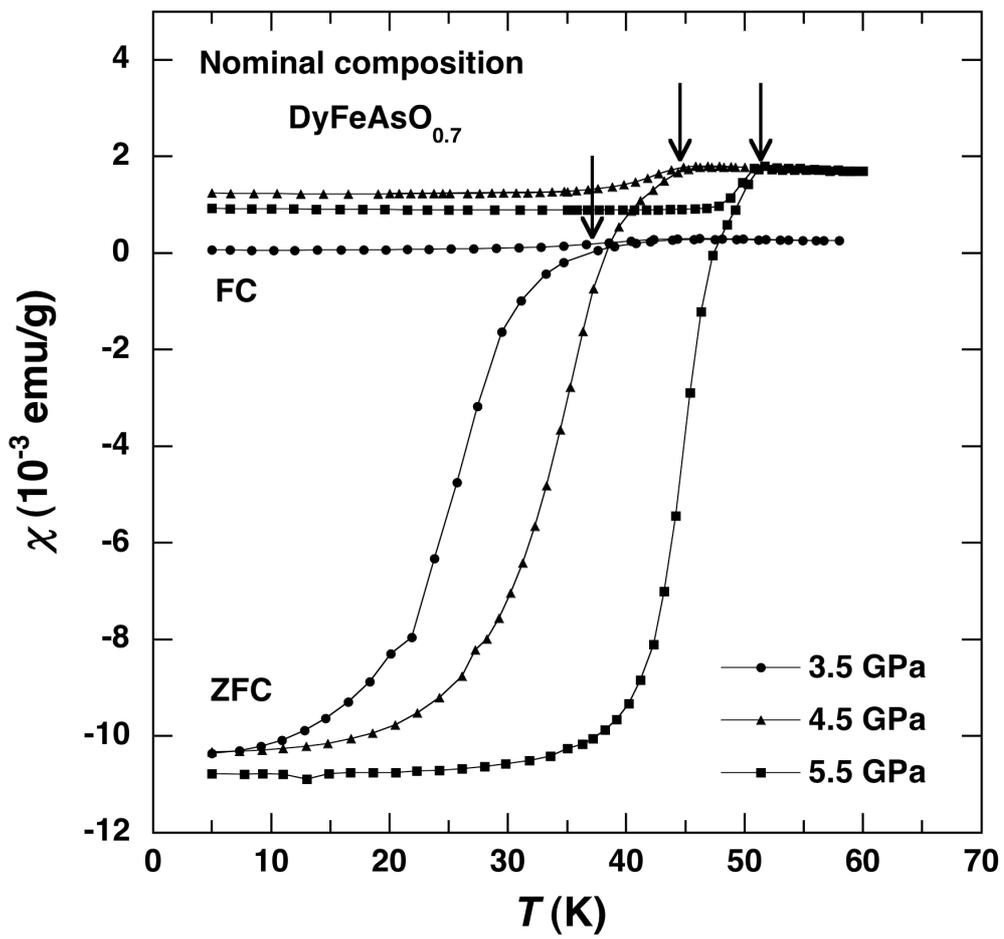

Fig. 3

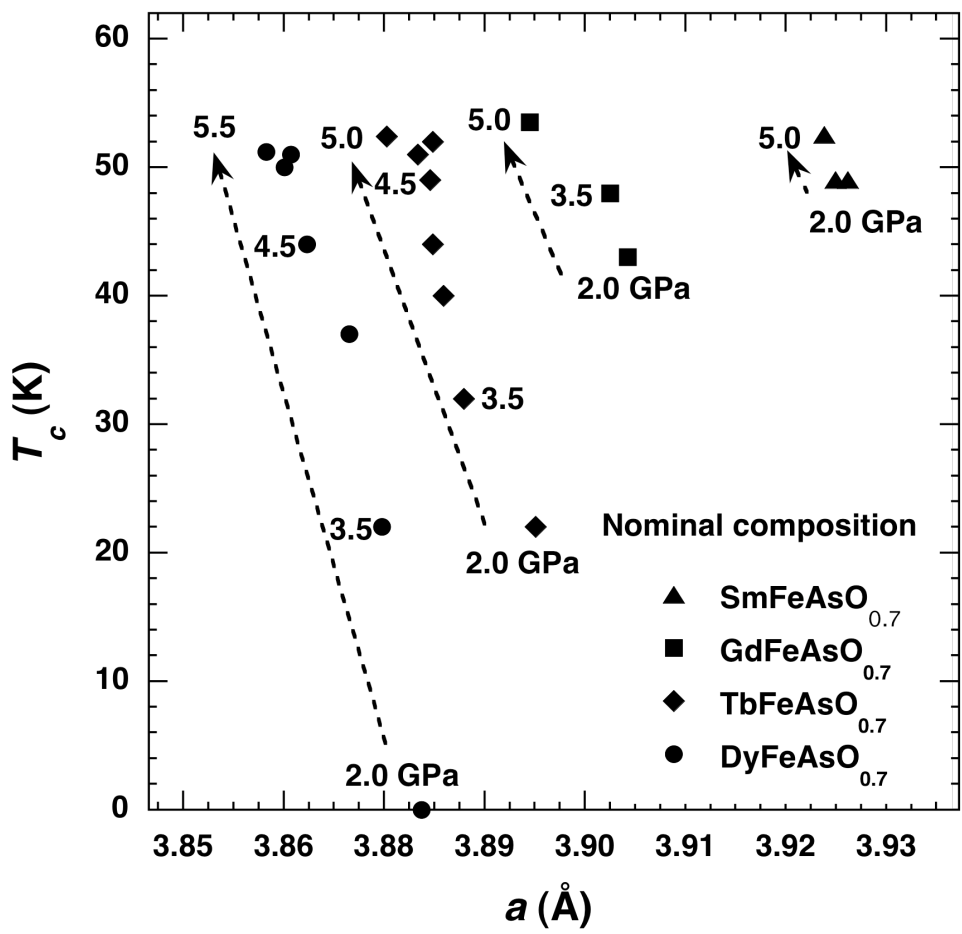

Fig. 4

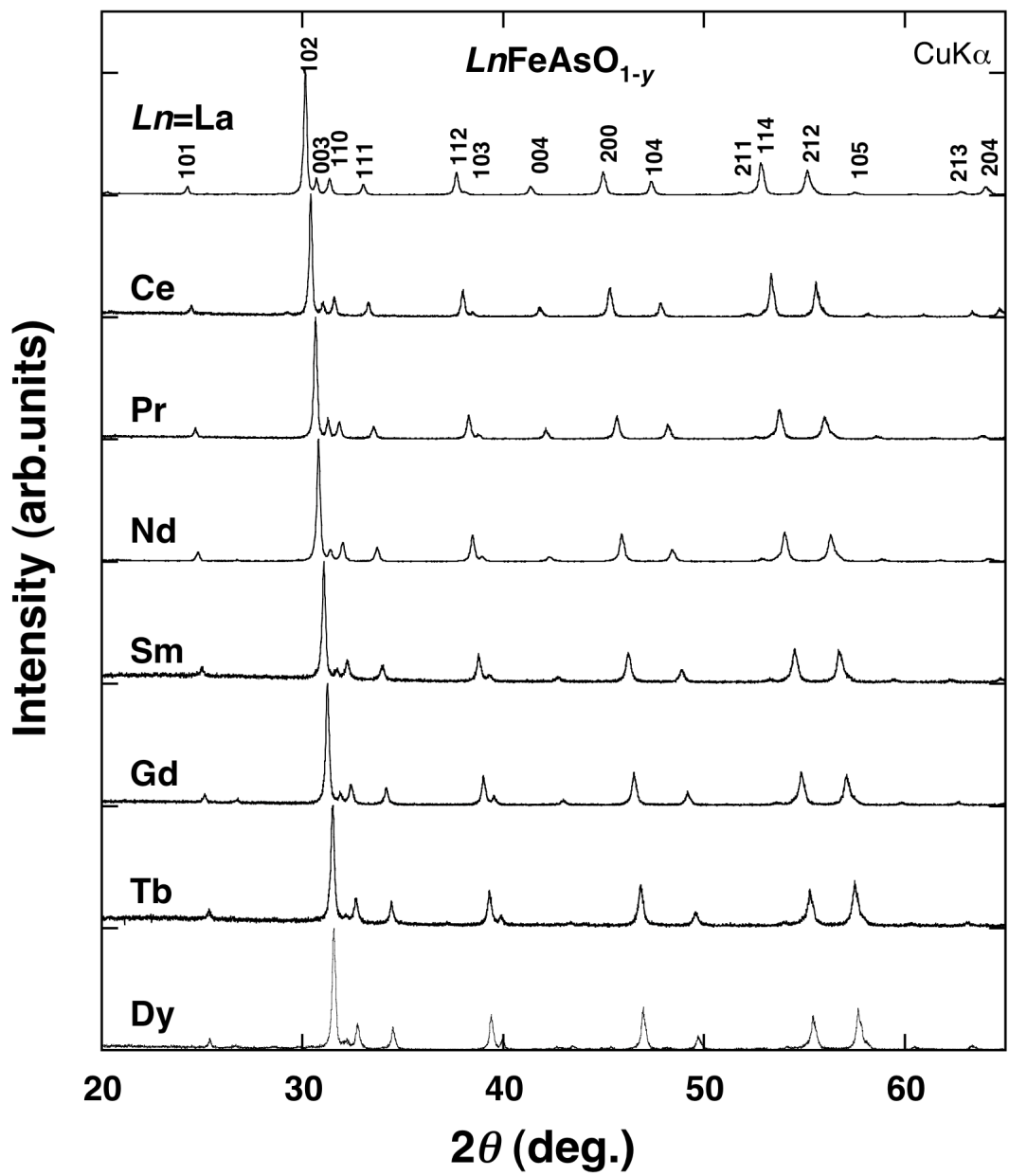

Fig. 5

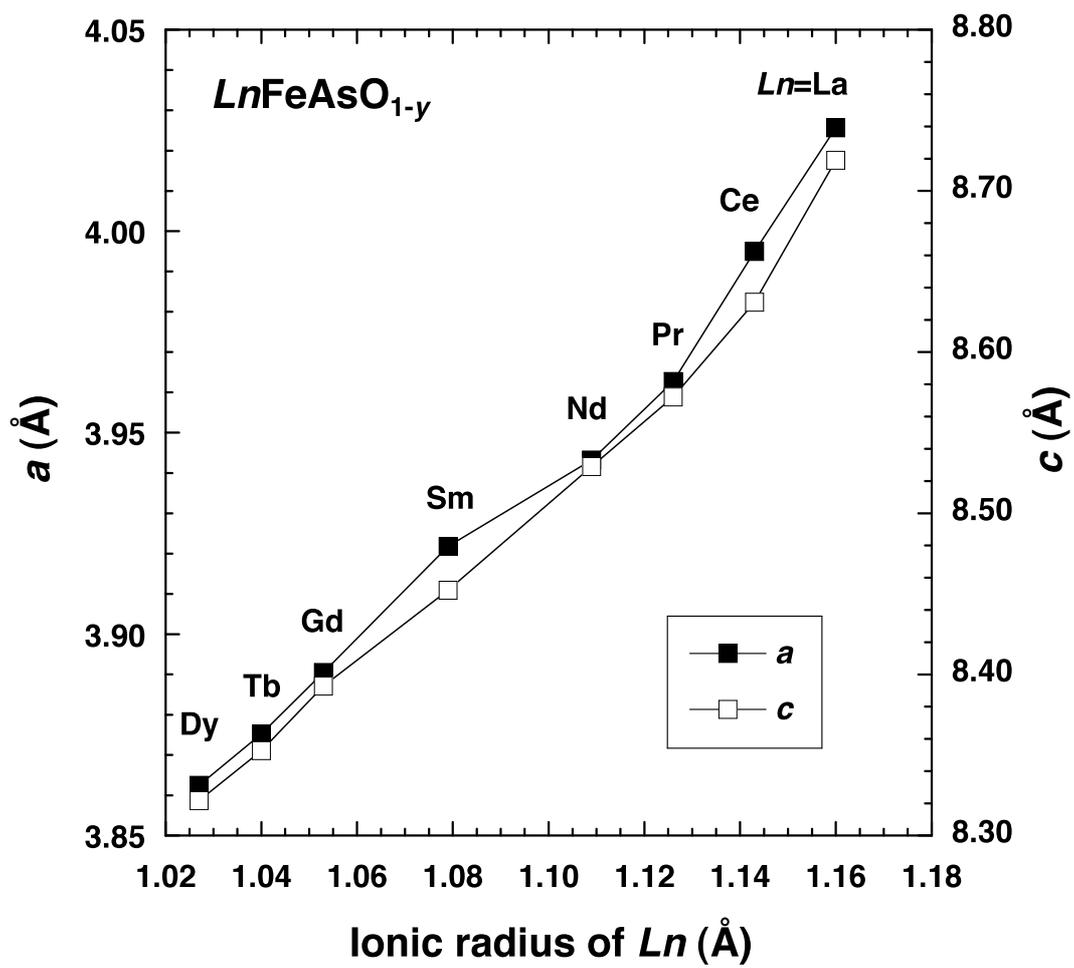

Fig. 6

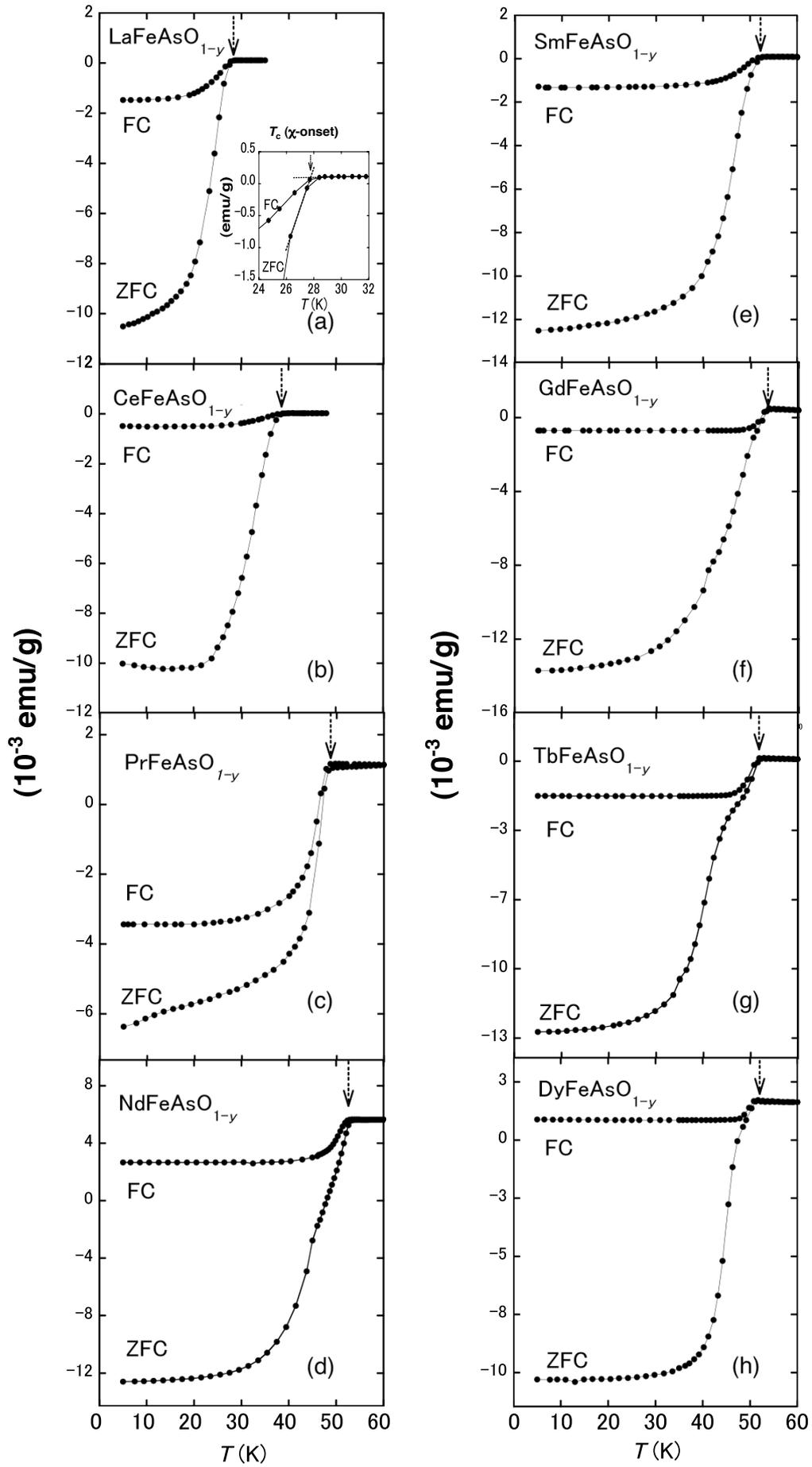

Fig. 7

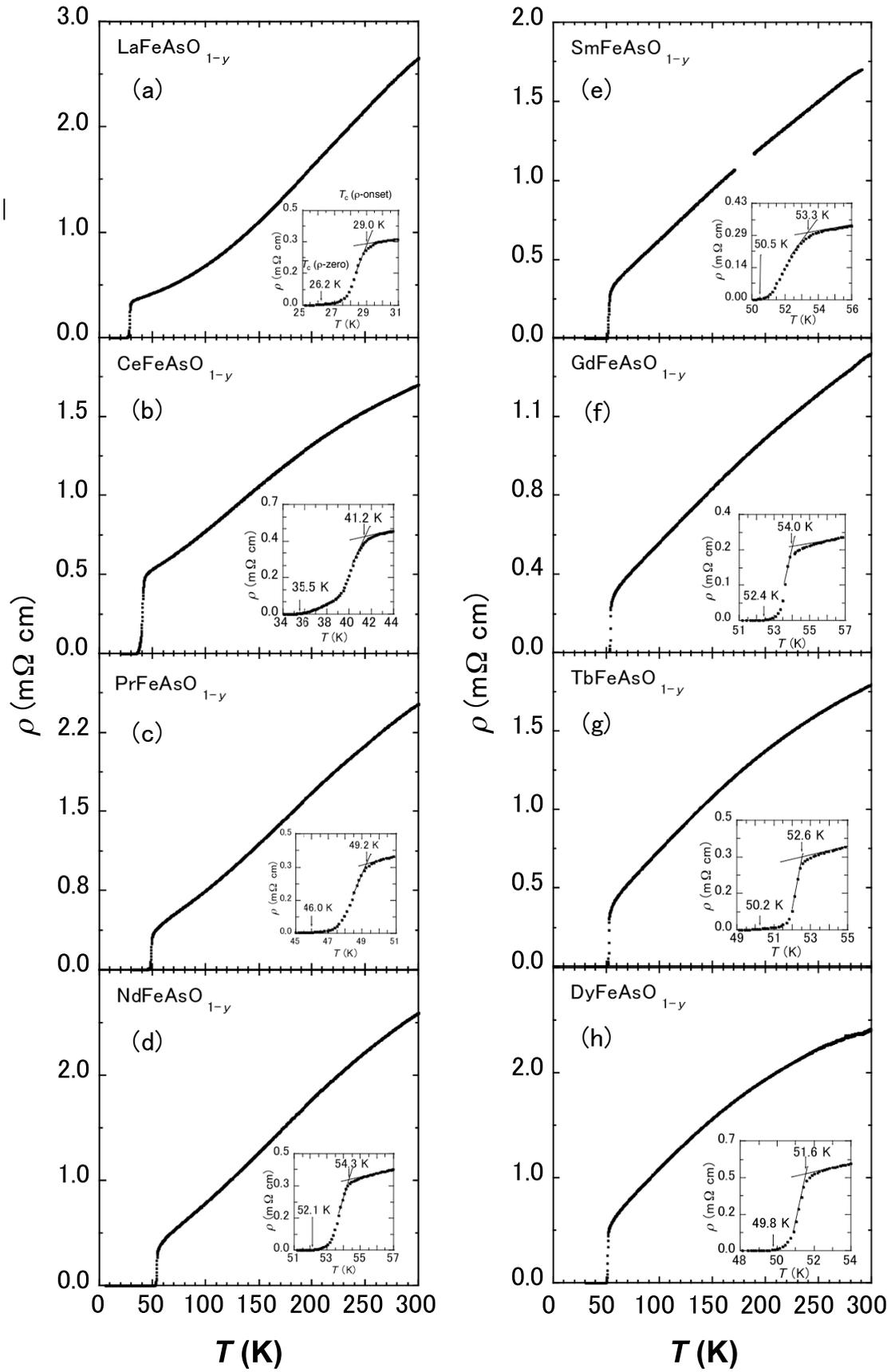

Fig. 8

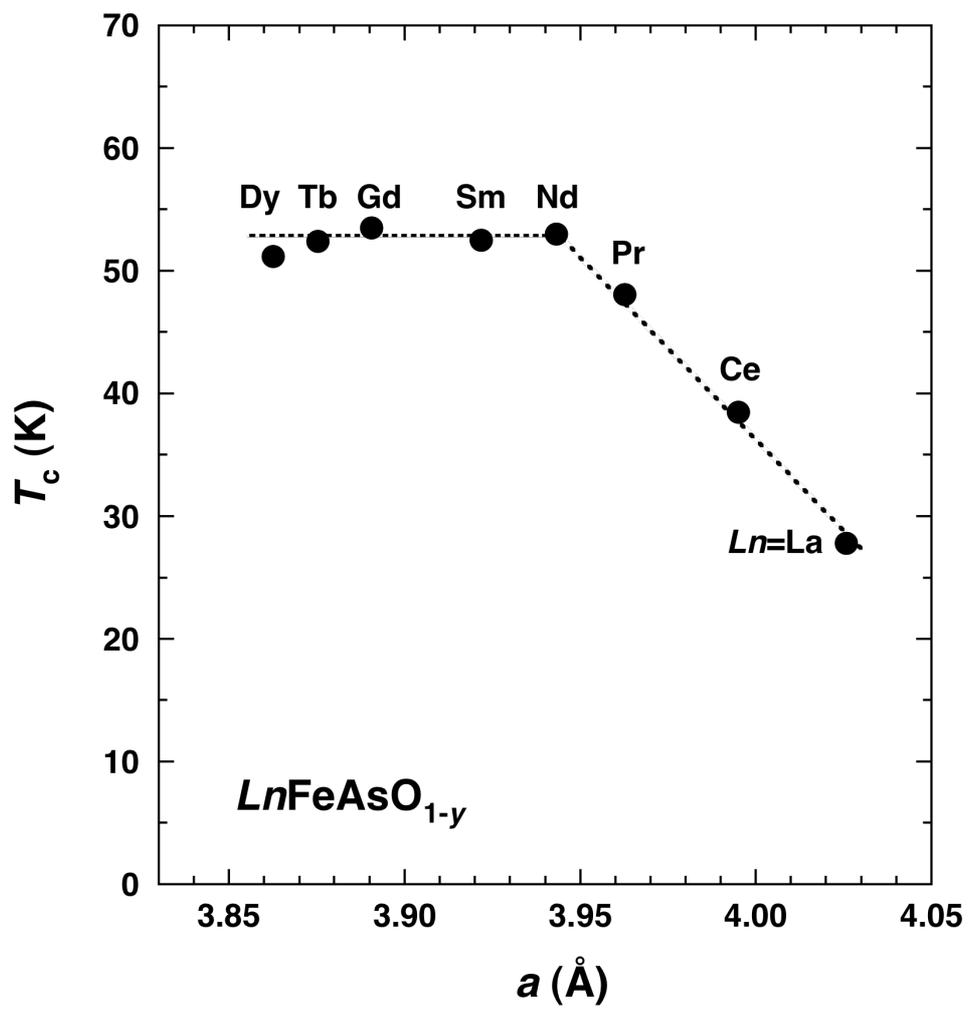

Fig. 9